\RequirePackage[svgnames]{xcolor}

\documentclass[11pt,letterpaper]{mystyle}

\definecolor{UHScarlet}{HTML}{C8102E}      
\definecolor{UHDarkRed}{HTML}{990000}      
\definecolor{UHLightRed}{HTML}{FBE5E8}     
\definecolor{UHGray}{HTML}{6D6E71}         
\definecolor{TinaCrimson}{HTML}{C8102E}    
\definecolor{YaleBlue}{HTML}{990000}       
\definecolor{CalGoldHex}{HTML}{FBE5E8}     

\tcbset{
    titlebox/.style={
        colback=UHLightRed,
        colframe=UHScarlet,
        boxrule=0.5mm,
        arc=2mm,
        auto outer arc,
        left=5mm,
        right=5mm,
        top=5mm,
        bottom=5mm,
    }
}

\usepackage[all]{hypcap}
\usepackage[svgnames]{xcolor}
\usepackage[comma,authoryear,compress]{natbib}
\usepackage{hyperref}
\hypersetup{
    colorlinks = true,
    citecolor = {UHDarkRed},
    linkcolor = {UHScarlet},
    urlcolor = {UHScarlet},
}

\usepackage{microtype}
\usepackage{graphicx}
\expandafter\def\csname ver@subfig.sty\endcsname{}
\usepackage{booktabs}
\usepackage{float}
\usepackage{bigstrut}

\usepackage{amsmath}
\usepackage{amssymb}
\usepackage{mathtools}
\usepackage{amsthm}
\usepackage{mathrsfs}
\usepackage{nicefrac}
\usepackage{dsfont}
\usepackage{enumitem}
\usepackage{subcaption}
\usepackage{cleveref}

\usepackage[utf8]{inputenc}
\usepackage[T1]{fontenc}
\usepackage{url}
\usepackage{amsfonts}
\usepackage{fdsymbol}
\usepackage{wrapfig}
\usepackage{lipsum}
\usepackage{stackengine}
\usepackage[font=small,labelfont=bf]{caption}
\usepackage{color}
\usepackage{adjustbox}
\usepackage{rotating}
\usepackage{makecell}
\usepackage{array}

\usepackage{tikz}
\usepackage{pgfplots}
\pgfplotsset{compat=1.16}

\definecolor{lightblue}{rgb}{0.22,0.45,0.70}
\definecolor{Gray}{gray}{0.95}
\definecolor{Cornsilk}{rgb}{1.0, 0.97, 0.86}

\graphicspath{{figures/}}

\title{When Agents Act on Web3: An Attack-Surface Survey of MCP, Skills, and Tool Calling}

\runningtitle{When Agents Act on Web3: An Attack-Surface Survey of MCP, Skills, and Tool Calling}

\author{
  Rabimba Karanjai$^{1,3}$,
  Yang Lu$^{1}$,
  Nour Diallo$^{1}$,
  Wujie Xiong$^{2}$,
  Lei Xu$^{2}$, and
  Weidong (Larry) Shi$^{1}$
}

\affil[1]{University of Houston, USA}
\affil[2]{Kent State University, USA}
\affil[3]{PayPal AI Labs, USA}

\correspondingauthor{Rabimba Karanjai, University of Houston and PayPal AI Labs}


\begin{document}

\begin{abstract}
AI agents increasingly act rather than merely read: across the Model Context Protocol (MCP) ecosystem, the share of deployed tools that modify external state has risen from 27\% to 65\% of tool use. When agents exercise this authority on public blockchains through MCP, skills, and tool calling, the consequences of an attack are governed by the blockchain execution layer rather than by conventional software assumptions. This survey argues that four properties of that layer (irreversibility, signing authority, continuous autonomy, and sequence-level composition) qualitatively change the threat model, turning the recoverable failures of generic agent security into a standing, irreversible loss. We organize the fragmented MCP-security literature into an attack-surface taxonomy, then contribute a Web3 risk-mapping matrix that ties each attack class to its amplified impact, the responsible amplifiers, a representative mitigation, and the residual gap. We synthesize defenses, including emerging blockchain-based mechanisms, and find them improving but insufficient: measured protections stop fewer than 30\% of attacks, and model-level safety refuses fewer than 3\%. We close by positioning the work against adjacent surveys and deriving a research agenda from the matrix's open cells.
\end{abstract}

\maketitle
\vspace{3mm}

\section{Introduction}\label{sec:intro}

AI agents have crossed a threshold from reading to acting. Across a sixteen-month window of the Model Context Protocol (MCP) ecosystem, the share of deployed agent tools that take \textit{actions}, modifying external state rather than only observing it, rose from 27\% to 65\% of tool use, while the number of distinct tools grew from roughly 5,000 to over 177,000 \cite{stein2026ecosystem}. Among the domains where this shift is fastest are financial ones, where an agent's action is a transaction.

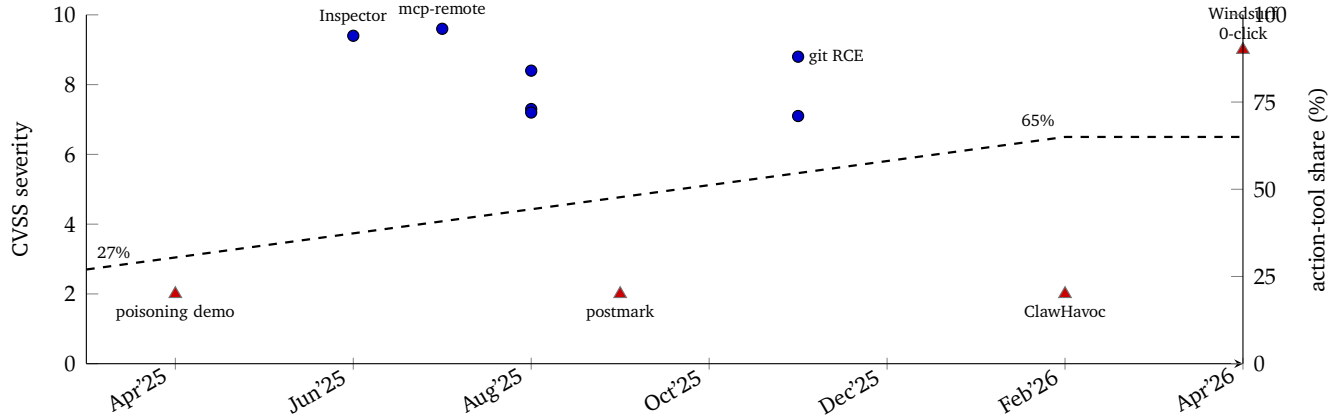
\begin{figure}[t]
\centering
\resizebox{\columnwidth}{!}{%
\begin{tikzpicture}
\begin{axis}[
  width=0.82\columnwidth, height=4.4cm,
  scale only axis,
  axis y line*=left,
  axis x line=bottom,
  xmin=0, xmax=13, ymin=0, ymax=10,
  xtick={1,3,5,7,9,11,13},
  xticklabels={Apr'25,Jun'25,Aug'25,Oct'25,Dec'25,Feb'26,Apr'26},
  x tick label style={font=\scriptsize, rotate=30, anchor=east},
  ylabel={\scriptsize CVSS severity},
  ytick={0,2,4,6,8,10}, y tick label style={font=\scriptsize},
  ylabel style={font=\scriptsize},
  clip=false,
]
\addplot+[only marks, mark=*, mark size=2pt, color=black] coordinates {
  (3,9.4)   
  (4,9.6)   
  (5,8.4)   
  (5,7.3)   
  (5,7.2)   
  (8,8.8)   
  (8,7.1)   
};
\addplot+[only marks, mark=triangle*, mark size=2.6pt, color=black!55] coordinates {
  (1,2.0)   
  (6,2.0)   
  (11,2.0)  
  (13,9.0)  
};
\node[font=\tiny,anchor=south] at (axis cs:3,9.4) {Inspector};
\node[font=\tiny,anchor=south] at (axis cs:4,9.6) {mcp-remote};
\node[font=\tiny,anchor=west] at (axis cs:8,8.8) {git RCE};
\node[font=\tiny,anchor=north,text width=1.5cm,align=center] at (axis cs:1,2.0) {poisoning demo};
\node[font=\tiny,anchor=north] at (axis cs:6,2.0) {postmark};
\node[font=\tiny,anchor=north] at (axis cs:11,2.0) {ClawHavoc};
\node[font=\tiny,anchor=south,text width=1.3cm,align=center] at (axis cs:13,9.0) {Windsurf 0-click};
\end{axis}
\begin{axis}[
  width=0.82\columnwidth, height=4.4cm, scale only axis,
  axis y line*=right, axis x line=none,
  xmin=0, xmax=13, ymin=0, ymax=100,
  ylabel={\scriptsize action-tool share (\%)},
  ytick={0,25,50,75,100}, y tick label style={font=\scriptsize},
  ylabel style={font=\scriptsize},
]
\addplot[dashed, thick, color=black] coordinates {(0,27) (11,65) (13,65)};
\node[font=\tiny, anchor=south west] at (axis cs:0,27) {27\%};
\node[font=\tiny, anchor=south east] at (axis cs:11,65) {65\%};
\end{axis}
\end{tikzpicture}
}
\caption{Escalation of confirmed MCP/tool-calling security disclosures (Apr 2025--Apr 2026) against ecosystem growth. Circles: CVEs (height = CVSS). Triangles: documented incidents. Dashed line: share of \emph{action} tools rising 27\%$\rightarrow$65\% as total tools grew 5{,}000$\rightarrow$177{,}436.}
\label{fig:timeline}
\end{figure}

MCP, introduced in late 2024 and since placed under open governance, has become the dominant standard for connecting agents to external tools \cite{mcp-spec,mcp-aaif}. Its rapid adoption has been matched by a rapid accumulation of security failures. Within roughly a year, confirmed vulnerabilities progressed from the first demonstration of tool poisoning to a cluster of high-severity remote code execution and authentication flaws, and to the first documented malicious MCP server (Figure~\ref{fig:timeline}).

These two trends (agents gaining authority to act, and the tools that grant that authority proving insecure) intersect most sharply on public blockchains. When an agent uses MCP, skills, or tool-calling to act on a Web3 system, the consequences of a successful attack are governed not by conventional software assumptions but by the blockchain execution layer. This paper's thesis is that four properties of that layer (\textbf{irreversibility}, \textbf{signing authority}, \textbf{continuous autonomy}, and \textbf{sequence-level composition}) fundamentally change the threat model, turning the recoverable failures of generic agent security into standing, irreversible loss.

We treat the \textit{usage-to-risk} direction as primary: our subject is the attack surface that opens when agents act on Web3, not the separate question of using blockchains to secure agents, which we address only as a class of mitigations.

This paper makes five contributions:

\begin{itemize}
\item A \textbf{web3 risk-mapping matrix} (Table~\ref{tab:matrix}) that ties each attack class to its amplified Web3 impact, the amplifiers responsible, a representative mitigation, and the residual gap,  exposing where current defenses are inadequate.
\item A \textbf{four-amplifier framework} that explains \textit{why} the blockchain execution layer changes the threat model, distinguishing Web3 agent risk from generic agent risk.
\item A \textbf{synthesis of defenses with honest coverage gaps}, including a moderate treatment of blockchain-based defenses, mapped to the attack surface.
\item An \textbf{attack-surface taxonomy} (Table~\ref{tab:taxonomy}) that organizes the fragmented MCP-security literature along surface and lifecycle-stage axes, serving as the scaffolding for the analysis above.
\item A \textbf{scoped positioning} against adjacent surveys and a \textbf{research agenda} drawn from the matrix's residual gaps.
\end{itemize}

We scope the survey to the agent-to-Web3 attack surface and its mitigation. The distinct problem of agents as autonomous \textit{generators} of exploits, where the agent is the attacker rather than the victim or conduit, is out of scope (Section~\ref{sec:related}) \cite{exploit-gen-a1}.

\section{Background}\label{sec:bg}

This section establishes the protocol mechanics, the surrounding agent stack, and the key-custody and identity primitives that the rest of the paper analyzes. Readers familiar with MCP may skip \ref{sec:tool_calling} -- \ref{sec:mcp_mechanics}. 

\subsection{Tool calling, MCP, and skills}
\label{sec:tool_calling}
Three mechanisms enable a language model to act beyond text generation, each with distinct trust profiles. \textit{Tool calling} (or function calling) lets a model invoke a developer-defined function through a structured call. It emerged from work on reasoning-and-acting agents and tool-augmented models \cite{yao2023reactsynergizingreasoningacting,schick2023toolformerlanguagemodelsteach,patil2023gorillalargelanguagemodel}. \textit{MCP} standardizes this across providers, defining a common interface so that any compliant client can use any compliant tool server \cite{hou2025landscape}. \textit{Skills} (or plugins) package tools, instructions, and resources into installable units. The trust profiles differ. A tool called trusts the developer who wrote the function. MCP additionally trusts whoever operates the server and authored its tool descriptions. A skill trusts whoever published the package. Each added layer of indirection is an additional party to trust and an additional attack surface, as shown later.

\subsection{MCP mechanics}
\label{sec:mcp_mechanics}
\begin{figure}[t]
\centering
\resizebox{\columnwidth}{!}{%
\begin{tikzpicture}[font=\scriptsize, >=latex]
\tikzset{box/.style={draw, rounded corners=2pt, minimum height=0.9cm, align=center, inner sep=3pt}}
\node[box, minimum width=1.7cm] (host) at (0,0) {Host\\(LLM agent)};
\node[box, minimum width=1.5cm] (client) at (2.7,0) {Client};
\node[box, minimum width=1.7cm, fill=black!6] (srv) at (5.6,0) {Server(s)\\(tools)};
\draw[<->, thick] (host) -- node[above,font=\tiny]{} (client);
\draw[<->, thick] (client) -- node[above,font=\tiny]{JSON-RPC} node[below,font=\tiny]{STDIO / HTTP} (srv);
\draw[dashed] (4.15,-1.1) -- (4.15,1.1);
\node[font=\tiny, anchor=south] at (4.15,1.1) {trust boundary};
\node[draw, dashed, fill=black!10, minimum width=1.9cm, align=center, font=\tiny] (desc) at (5.6,-1.7) {tool \emph{descriptions}\\(planning)};
\node[draw, dashed, fill=black!10, minimum width=1.9cm, align=center, font=\tiny] (out) at (2.7,-1.7) {tool \emph{outputs}\\(response handling)};
\draw[->, dotted] (desc) -- (srv);
\draw[->, dotted] (out) -- (client);
\node[font=\tiny, anchor=north] at (0,-0.55) {client-host surface};
\node[font=\tiny, anchor=north] at (5.6,-0.55) {server surface};
\end{tikzpicture}}
\caption{MCP trust boundaries and the two untrusted-content entry points. Tool descriptions enter at planning and tool outputs at response handling. Both cross into the agent's context as trusted input: ``the trust boundary is the schema'' (Sec.~\ref{sec:bg}). The four surfaces of Table~\ref{tab:taxonomy} map onto this dataflow.}
\label{fig:trust}
\end{figure}
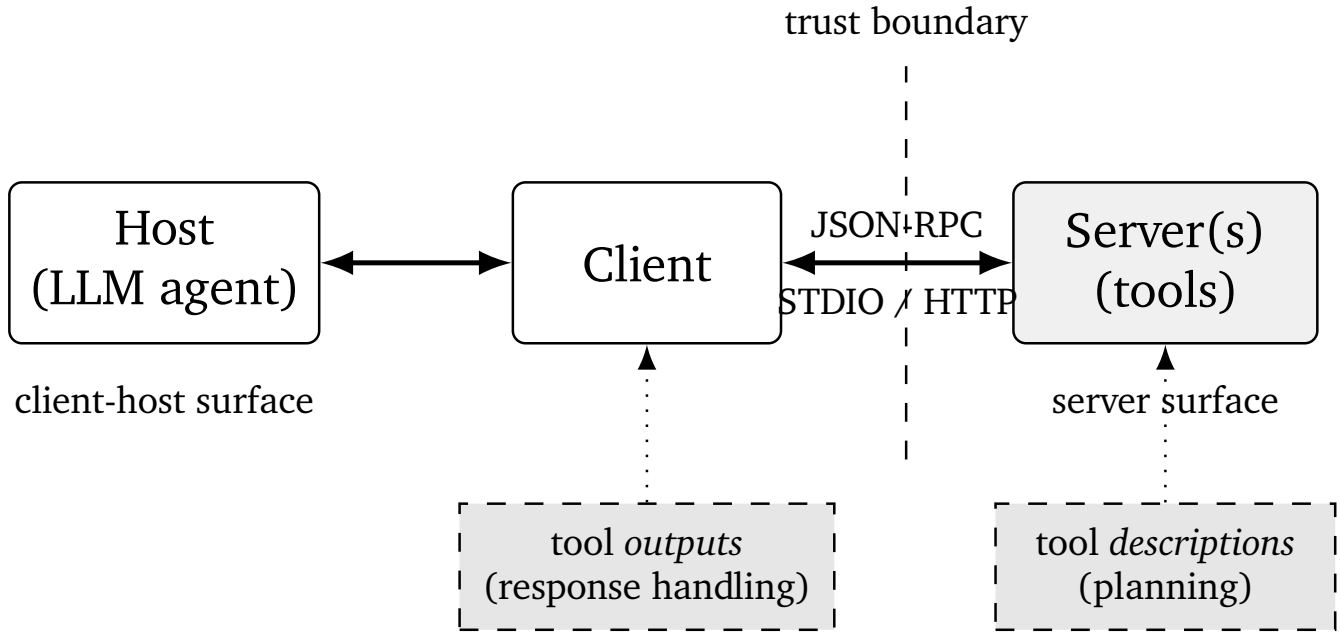

MCP follows a ``client host server'' model over JSON-RPC \cite{mcp-spec}. A host application embeds a client, which connects to one or more servers that expose tools. Client and server negotiate capabilities at connection time, and the host's model selects and invokes tools over the session. The protocol exposes several primitives (e.g., resources, prompts, and tools, with later additions for sampling and elicitation) and supports both local (STDIO) and networked (streamable HTTP) transports \cite{hou2025landscape}. The most security-critical mechanism here is that a tool's \textit{description}, the natural-language metadata that declares what the tool does and how to call it, is supplied to the model as trusted context at planning time. The model has no native means to distinguish a description authored by an honest server from one authored by an adversary. In MCP, the trust boundary is the schema. This single design fact seeds the largest class of protocol-level attacks in Section~\ref{sec:taxonomy}. Figure~\ref{fig:trust} makes the boundary concrete.

\subsection{The agent stack and Web3}\label{sec:bg-stack}

\begin{figure}[t]
\centering
\begin{tikzpicture}[font=\scriptsize, node distance=0pt]
\tikzset{layer/.style={draw, minimum width=7.2cm, minimum height=0.62cm, anchor=south}}
\node[layer] (mcp) at (0,0) {MCP — tool access (Resources / Prompts / Tools)};
\node[layer, fill=black!4] (a2a) at (0,0.62) {A2A — agent-to-agent messaging};
\node[layer, fill=black!8] (ap2) at (0,1.24) {AP2 — payment authorization};
\node[layer, fill=black!12] (x402) at (0,1.86) {x402 — stablecoin settlement (HTTP 402)};
\node[anchor=south, font=\scriptsize\bfseries] at (0,2.55) {Agent stack};
\begin{scope}[yshift=-1.7cm]
\draw[->, thick] (-3.6,0) -- (3.6,0);
\foreach \x/\lab in {-3.2/{raw\\key}, -1.6/{session\\keys}, 0/{EIP-7702\\delegation}, 1.6/{MPC\\split}, 3.2/{TEE\\custody}}{
  \filldraw (\x,0) circle (1.6pt);
  \node[align=center, anchor=north, text width=1.3cm] at (\x,-0.12) {\lab};
}
\node[anchor=south west, font=\scriptsize] at (-3.6,0.08) {autonomy $\uparrow$};
\node[anchor=south east, font=\scriptsize] at (3.6,0.08) {human control $\uparrow$};
\node[anchor=south, font=\scriptsize\bfseries] at (0,0.6) {Signing-authority custody spectrum};
\end{scope}
\end{tikzpicture}
\caption{The agent stack (top) and the signing-authority custody spectrum (bottom). MCP at the tool layer is where the attacks of Table~\ref{tab:taxonomy} land. The custody spectrum, from raw keys to TEE-isolated keys, governs the blast radius of the signing-authority amplifier.}
\label{fig:stack}
\end{figure}
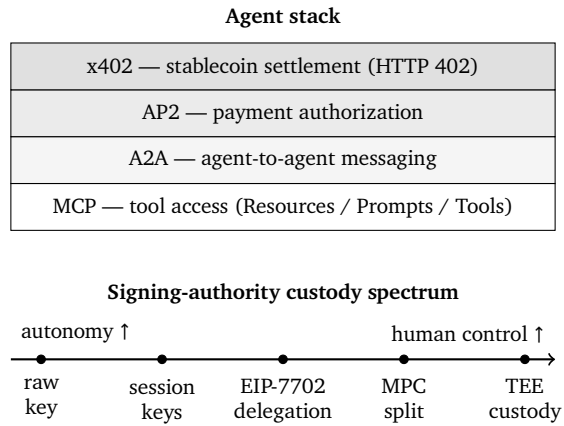

MCP rarely operates alone. A surrounding stack has emerged. The Agent2Agent (A2A) protocol carries messages between agents \cite{a2a}. The Agent Payments Protocol (AP2) standardizes the process by which an agent obtains authorization to pay \cite{ap2}. And x402 revives the HTTP 402 status code to settle stablecoin payments over a web request \cite{x402}. The upper half of Figure~\ref{fig:stack} shows the layering: MCP grants tool access, A2A carries inter-agent messages, AP2 authorizes payment, and x402 settles it. The stack matters for this survey because it is where an agent acquires \textit{payment and signing authority}, the capability governed by the next subsection's custody models.

\subsection{Agent-with-wallet key custody}\label{sec:bg-custody}

When an agent transacts on a blockchain, the central design question is how it holds signing authority. The options form a spectrum from least to most constrained (lower half of Figure~\ref{fig:stack}). At one extreme, an agent holds a \textit{raw private key}, granting unrestricted signing: operationally simple and maximally dangerous. \textit{Session keys} limit a key's scope or lifetime. EIP-7702, shipped in Ethereum's 2025 Pectra upgrade, lets an externally owned account delegate to contract code, making the scope of an agent's signing authority programmable. If the delegation is broad, then it will be broadly dangerous \cite{eip7702}. \textit{Multi-party computation (MPC)} splits signing authority so no single party (including the agent) can sign alone \cite{autonomous-agents-blockchain,staley2026reference,session-mpc-custody}. \textit{Trusted execution environments (TEEs)} isolate keys in attested hardware \cite{M_n_trey_2022,shepherd2018remote}. We define these here once. Section~\ref{sec:usage} treats them as deployment patterns and Section~\ref{sec:def} as defenses.

\subsection{Agent identity and authorization}\label{sec:bg-identity}

Distinct from \textit{what an agent can sign} is \textit{who an agent is} and \textit{what it is permitted to do}. Access to MCP servers is increasingly mediated by OAuth 2.1. Inter-agent identity is asserted through signed agent cards or decentralized identifiers, and transport-level authentication uses mutual TLS \cite{agent-auth-oauth,a2a-security,chang2025agentnetworkprotocoltechnical,w3c2022didcore,owasp}. These matter to this survey for a specific reason. Identity is simultaneously an \textit{attack surface} (unsigned or replayable identity metadata enables impersonation (Section~\ref{sec:taxonomy})) and a \textit{defense} (signed cards and mutual TLS are among the mitigations of Section~\ref{sec:def}). 

\section{How Agents Use MCP on Web3}\label{sec:usage}

Section~\ref{sec:bg} described the primitives. This section describes how they are actually deployed, and the design patterns and anti-patterns that shape the resulting attack surface.

\subsection{Integration modes}

The first thing that shapes the attack surface is how deeply an agent is wired into the chain. Agent interaction with a blockchain falls into three modes. On the \textit{read path}, the agent queries on-chain state (e.g., balances, prices, contract state) through RPC-backed tools. This is the lowest-risk mode, as no signing is involved. On the \textit{write path}, the agent builds, signs, and submits transactions. This is where the signing authority (Section~\ref{sec:bg-custody}) is exercised and where the consequences of a compromised tool call become irreversible. The third mode is \textit{machine-to-machine payment}, in which agents pay for resources or each other's services through x402-style settlement \cite{x402}. Production MCP servers already span all three modes, from general multi-chain servers to exchange-operated trading integrations \cite{web3-mcp,evm-mcp,onchain-mcp,bitcoin-mcp,bybit-trading-mcp}.

\subsection{Design patterns}

Several recurring patterns shape how agents are deployed against Web3, and some double as mitigation. A \textit{stateless per-request server} holds no signing authority between calls, narrowing the window of exposure. \textit{Multi-chain tool-gating} restricts which chains and which operations a given tool may touch, an application of least privilege that also limits blast radius. \textit{Split-authority custody} (MPC) and \textit{policy-constrained wallets} (wallets that enforce per-transaction limits or allow-lists) keep an agent from unilaterally moving arbitrary value. We note these here as deployment patterns. Section~\ref{sec:def} returns to the gateway/proxy pattern as a defense in its own right.

\subsection{Anti-patterns}

The corresponding anti-patterns recur across deployments and map directly to the attack classes of Section~\ref{sec:taxonomy}. \textit{Over-permissioning} (e.g., granting an agent broad, standing signing authority rather than scoped, revocable authority) is the most consequential. An industry analysis of MCP implementations found injection-class weaknesses in a large fraction of deployments \cite{endor-labs}.  The \textit{sanitization gap} (trusting tool inputs and outputs on the local STDIO channel without validation) underlies the command-injection class. And \textit{plugin-as-trust-decision} (installing a skill or server without treating it as the security decision it is) is what makes supply-chain compromise effective. The same marketplace dynamics that grew one skill registry past 10,700 entries also carried hundreds of malicious skills \cite{skills-clawhub}.

\section{An Attack-Surface Taxonomy for Agent Tool-Use}\label{sec:taxonomy}

The security literature on the Model Context Protocol has grown quickly but unevenly: individual disclosures, benchmark papers, and vendor analyses each describe attacks against a different part of the protocol, using incompatible vocabularies. To reason about which of these attacks matter when an agent acts on a blockchain, they must first be organized. Throughout this paper, we assume an adversary who can author, operate, interpose on, or plant content reachable by the agent, but who cannot break standard cryptographic primitives or read keys in properly isolated custody (MPC or TEE). The adversary's goal is to induce an action favoring the attacker, such as an unauthorized signed transaction or a leak of sensitive context. The agent is thus the victim or unwitting conduit, never the attacker, with the inverse case (the agent generating exploits) a distinct model excluded in Section~\ref{sec:related}. This section adopts two axes from prior systematization as organizing scaffolding and uses them to locate each known attack class. The Web3-specific analysis enabled by this scaffolding is deferred to Section~\ref{sec:amplifiers}.

\subsection{Organizing axes}

We organize the attack surface along two axes. The first, the \textit{surface} axis, follows MCPSecBench \cite{mcpsecbench} in distinguishing where an attack enters: the \textbf{user} (the human-supplied request and approvals), the \textbf{client-host} (the agent runtime and its model), the \textbf{transport} (the JSON-RPC channel between client and server), and the \textbf{server} (the tool provider). The second, the \textit{lifecycle-stage} axis, follows MSB \cite{msb} in distinguishing when an attack acts within the MCP tool-use pipeline: \textbf{task planning} (when the agent decides which tool to call), \textbf{tool invocation} (when the call executes), and \textbf{response handling} (when the tool's output re-enters the agent's context).

Cutting across both axes is a distinction between \textbf{protocol-level} weaknesses, which arise from the MCP design itself (most importantly, its treatment of tool descriptions and tool outputs as trusted context)  and \textbf{implementation-level} weaknesses, which arise from how a particular server or client is built.  The distinction matters because the two classes demand different defenses: a protocol-level weakness recurs across every conforming implementation, whereas an implementation-level weakness can be patched in a single codebase.  We map the resulting categories to the established OWASP and MAESTRO agent-threat vocabularies \cite{breaking-protocol} where they align, so that the taxonomy can be read alongside existing practitioner frameworks \cite{owasp-maestro}.

Table~\ref{tab:taxonomy} presents the taxonomy. Each row is an attack class, located on both axes and labeled protocol- or implementation-level, with a representative confirmed vulnerability or documented incident. We populate the representative column only with vulnerabilities that carry a confirmed identifier or a publicly documented incident, so that the taxonomy is anchored to demonstrated rather than hypothetical attacks.

\begin{table}[t]
\caption{Attack-Surface Taxonomy for Agent Tool-Use Against Web3. Axes adopted from MCPSecBench (surface) and MSB (stage) as organizing scaffolding. Each class carries a representative confirmed CVE or documented incident.}
\label{tab:taxonomy}
\centering
\footnotesize
\begin{tabular}{@{}p{0.5cm}p{3.0cm}p{1.8cm}p{2.0cm}p{0.7cm}p{5.5cm}@{}}
\toprule
\# & Attack class & Surface & Stage & Layer & Representative CVE / incident (verified) \\
\midrule
1 & Tool poisoning & server & task planning & P & CVE-2025-54136 \textit{MCPoison} (CVSS 7.2); MCPTox 72.8\% ASR \\
2 & Indirect prompt injection & client-host & response handling & P & CVE-2025-54135 \textit{CurXecute} \\
3 & Command / STDIO injection & transport & tool invocation & I & CVE-2025-6514 \textit{mcp-remote} (9.6); CVE-2026-30615 \textit{Windsurf} \\
4 & RCE via reference-impl. flaw & server & tool invocation & I & CVE-2025-68143/68144 \textit{mcp-server-git} (8.8/7.1) \\
5 & Sandbox / path-containment escape & server & tool invocation & I & CVE-2025-53109/53110 \textit{EscapeRoute} (8.4/7.3) \\
6 & Auth / authorization gap & client-host & tool invocation & I & CVE-2025-49596 \textit{MCP Inspector} (9.4) \\
7 & Supply-chain (malicious server/skill) & server & task planning & I & \textit{postmark-mcp} (first documented); 341$\rightarrow$824 malicious skills \\
8 & Preference manipulation & server & task planning & P & MPMA preference-manipulation \\
9 & Identity / impersonation & transport & task planning & P & A2A Agent-Card tampering / replay \\
\bottomrule
\end{tabular}
\end{table}

\subsection{Attack classes}

We describe each class briefly. Table~\ref{tab:taxonomy} gives the full mapping.

\textbf{Tool poisoning.} The defining protocol-level weakness of MCP is that a tool's natural-language description is injected into the agent's context and treated as trusted. An adversary who controls a tool description can therefore embed instructions for the model to follow at planning time. The attack was first demonstrated by Invariant Labs \cite{invariant-toolpoison} and has since been shown to be highly effective at scale. The MCPTox benchmark reports a 72.8\% attack-success rate against a leading model, with refusal rates below 3\% even for safety-tuned models \cite{mcptox}, indicating that model alignment alone does not mitigate the attack. A configuration-swap variant against a popular MCP client was assigned CVE-2025-54136 (CVSS 7.2) \cite{mcpoison}.

\textbf{Indirect prompt injection.} Distinct from a poisoned tool description, indirect prompt injection arrives in the \textit{content} an agent retrieves (e.g., a web page, a file, an email) and is acted upon when that content re-enters the model's context at response handling. The general class predates MCP \cite{greshake-ipi}, but MCP broadens it by making external content trivially reachable through tools. A documented instance against a popular agentic IDE was assigned CVE-2025-54135 \cite{curxecute}.

\textbf{Command and STDIO injection.} At the transport layer, MCP servers that pass arguments to a shell or that trust the local STDIO channel are exposed to command injection. The most severe confirmed example, in the widely used `mcp-remote` connector, was assigned CVE-2025-6514 with a CVSS score of 9.6 \cite{mcp-remote-rce}. A zero-click prompt-injection-to-code-execution chain in another agentic IDE was assigned CVE-2026-30615 \cite{ox-stdio}.

\textbf{Remote code execution via reference-implementation flaws.} Several of the official MCP reference servers shipped exploitable code. The Git server carried an arbitrary-path and argument-injection pair, CVE-2025-68143 (CVSS 8.8) and CVE-2025-68144 (7.1) \cite{mcp-git-rce}. These are implementation-level. They were fixed in the affected servers, but their prevalence in \textit{reference} code meant the flaws propagated into many downstream deployments.

\textbf{Sandbox and path-containment escape.} The File system reference server permitted symlink- and containment-based escapes from its intended directory, assigned CVE-2025-53109 (CVSS 8.4) and CVE-2025-53110 (7.3) \cite{mcp-filesystem-escape}. The recurrence of containment bypasses across implementations suggests the sandboxing model, not any single bug, is the weak point.

\textbf{Authentication and authorization gaps.} Tooling around MCP frequently ships open by default. The MCP Inspector debugging tool exposed an unauthenticated interface allowing remote code execution, assigned CVE-2025-49596 with a CVSS score of 9.4 \cite{mcp-inspector-auth}.

\textbf{Supply-chain compromise.} Because MCP servers and agent skills are distributed through open registries with low barriers to publication, they inherit the supply-chain risks of any package ecosystem. The first publicly documented malicious MCP server, a counterfeit of a legitimate email connector, added a single line that blind-copied every sent message to an attacker-controlled address. It was downloaded approximately 1,500 times before removal \cite{postmark-backdoor}. At the skill layer, a coordinated campaign saw the count of malicious skills in one marketplace rise from 341 to 824 as the marketplace itself grew past 10,700 entries \cite{skills-clawhub}.

\textbf{Preference manipulation.} Beyond injecting instructions, an adversary can manipulate \textit{which} tool the agent prefers, steering it toward an attacker-controlled tool through crafted metadata. This preference-manipulation attack operates at planning time and largely evades per-call output checks \cite{mpma}.

\textbf{Identity and impersonation.}  At the multi-agent layer of the stack (Section~\ref{sec:bg-stack}), an agent's identity is asserted through metadata such as agent cards. When unsigned, they can be tampered with or replayed to impersonate a trusted agent. This surface is documented in analyses of the A2A layer \cite{a2a-security} and becomes consequential in Web3 precisely because an impersonated agent may hold signing authority.

Across all classes, an independent measurement of prevalence is provided by an industry analysis of 2,614 MCP implementations, which reported path-traversal, code-injection, and command-injection weaknesses in 82\%, 67\%, and 34\% of implementations, respectively \cite{endor-labs}.

\section{Why Web3 Changes the Threat Model: Four Amplifiers and a Risk Map}\label{sec:amplifiers}

The taxonomy of Section~\ref{sec:taxonomy} is not specific to blockchains. The same attack classes apply wherever agents call tools. What changes on Web3 is not the attacks but their \textit{consequences}. This section argues that four properties of the blockchain execution layer systematically amplify the impact of agent tool-use attacks, turning recoverable failures into irreversible loss, and then maps each attack class to its amplified Web3 impact in a risk-mapping matrix (Table~\ref{tab:matrix}), the paper's central contribution.

\subsection{Four amplifiers}

We identify four amplifiers. Each is a property of the Web3 setting, not of MCP, and each independently worsens the consequence of an attack that Section~\ref{sec:taxonomy} catalogs.

\textbf{Irreversibility.} A settled blockchain transaction cannot be reversed, charged back, or undone by an intermediary. Finality is the defining property of the settlement layer. In conventional software, an attacker who induces a wrong action can often be remediated: a transfer reversed, a session revoked, a backup restored. On-chain, the same induced action is permanent. Irreversibility is therefore the amplifier that converts every other impact in the matrix from a recoverable error into a standing loss. It is the unifying claim of this section.

\textbf{Signing authority.} An agent that holds, or can invoke, signing authority can move value directly: a single signed transaction is itself the harmful act, with no further step required. The progression of key-custody models in Section~\ref{sec:bg-custody} (from raw keys through scoped delegation under EIP-7702 to split-authority and hardware-isolated custody) exists precisely because granting an automated party signing authority is dangerous. EIP-7702's delegation semantics make the scope of that authority explicit and, when misconfigured, broad \cite{eip7702}. The amplifier is that in Web3, the blast radius of a compromised agent is financial by default.

\textbf{Continuous autonomy.} Agents operate continuously, ingest untrusted external content, and increasingly act without per-step human review. This is the amplifier behind the action-tool shift noted in Section~\ref{sec:intro}: the rise from 27\% to 65\% of tool use \cite{stein2026ecosystem}. The window between a successful injection and an irreversible on-chain action is bounded only by the agent's own latency, not by a human's.

\textbf{Sequence-level composition.} The fourth amplifier is the most subtle. Individually authorized tool calls can be composed into a sequence whose aggregate effect is malicious (e.g., a permitted read, a permitted computation, and a permitted transfer chaining into an exfiltration or drain pipeline) even though no single call violates a policy. Multi-turn evaluation shows that agents are substantially more vulnerable when attacks span several interactions than when judged call-by-call \cite{mcp-safetybench}, and documented sequence-level chains (authorized tool calls that combined into exfiltration pipelines) appear in incident analysis \cite{ox-stdio}. Static, per-call defenses are, by construction, blind to the composition.

These amplifiers are not independent in effect: irreversibility sets the stakes, signing authority and continuous autonomy determine how fast and how far an attack propagates, and sequence-level composition determines how easily it evades per-call defenses. Their conjunction is what distinguishes Web3 agent risk from generic agent risk.

\subsection{Mechanism}\label{sec:mechanism}

The cleanest demonstrated mechanism connecting an agent tool call to direct value movement is the malicious-router study, which examined intermediaries sitting between an agent and its model. Of 428 routers tested, 26 were found to inject malicious tool calls or access planted credentials, and the authors show that a single rewritten tool call is sufficient for arbitrary code execution \cite{inc-router-drain}. The authors further report a client-wallet drain of roughly half a million dollars. We note this figure with caution, as the reproducible demonstration involved minimal funding and the amount has not been independently confirmed. The mechanism (one rewritten call, executed with signing authority, settled irreversibly) is what matters here, and it holds regardless of the disputed figure.

\subsection{The risk-mapping matrix}

Table~\ref{tab:matrix} maps each attack class from the taxonomy to its dominant Web3 impact, the amplifier or amplifiers that magnify it, a representative mitigation (developed in Section~\ref{sec:def}), and the residual gap that the mitigation does not close.

\begin{table}[t]
\caption{Web3 Risk-Mapping Matrix. Amplifiers: Irr=irreversibility, Sig=signing authority, Aut=continuous autonomy, Seq=sequence-level composition. Blank/partial mitigations denote open problems (Sec.~\ref{sec:related}).}
\label{tab:matrix}
\centering
\footnotesize
\begin{tabular}{@{}p{0.4cm}p{2.6cm}p{2.4cm}p{1.4cm}p{0.5cm}p{2.6cm}p{3.2cm}@{}}
\toprule
\# & Attack class & Web3 impact & Amplifiers & Sev. & Mitigation (Sec.~\ref{sec:def}) & Residual gap \\
\midrule
1 & Tool poisoning & unauthorized tx; irreversible loss & Irr, Sig, Seq & H & Signed/immutable manifests (ETDI) & semantic poisoning of signed tool; refusal $<$3\% \\
2 & Indirect prompt injection & unauthorized tx; data exfil & Aut, Seq & H & Gateway inspection; HITL on writes & agent acts on consumed content by design \\
3 & Command / STDIO injection & irreversible loss; key compromise & Sig & H & Transport hardening; sandboxing & impl. coverage uneven; protections $<$30\% \\
4 & RCE via ref-impl flaw & key compromise; irreversible loss & Sig & H & Patching; least-privilege exec & zero-day window vs 24/7 autonomy \\
5 & Sandbox escape & data exfil; key compromise & Sig & M--H & Hardened sandbox; path-allowlist & containment bypasses recur \\
6 & Auth gap & unauthorized tx; key compromise & Sig & H & OAuth 2.1; mTLS; no default-open & misconfiguration prevalent \\
7 & Supply-chain & data exfil; irreversible loss & Aut, Sig & H & Registry provenance; signing & low publish barrier; trust-on-first-use \\
8 & Preference manipulation & unauthorized tx & Seq, Aut & M & Tool-selection policy; cross-check & subtle steering evades per-call checks \\
9 & Identity / impersonation & unauthorized tx & Sig, Seq & M & Signed Agent Cards; mTLS & delegation-chain identity unsolved \\
\bottomrule
\end{tabular}
\end{table}

Three observations follow from the matrix. First, the impact of every row is qualitatively worsened by irreversibility. The same tool-poisoning or injection attack that yields a recoverable error off-chain yields a standing loss on-chain. Second, sequence-level composition recurs across the tool-poisoning, indirect-injection, supply-chain, and preference-manipulation rows. It is the amplifier least addressed by current defenses, because those defenses operate per call. Third, and most consequential for the research agenda, several rows have no adequate mitigation measures. The matrix's blank and partial mitigation cells are not omissions but findings, and they define the open problems of Section~\ref{sec:related}.

\section{Defenses and Their Limits}\label{sec:def}

The mitigation column of Table~\ref{tab:matrix} points to the defenses surveyed here. We organize them as families mapped to the attack surface, treat blockchain-based defenses at moderate depth, and close with an honest account of what current defenses do not cover.

\subsection{Defense families}

\textbf{Defense-in-depth and gateways.} The most widely advocated posture is layered defense around the MCP boundary \cite{mcp-guard,narajala-enterprise}. Its most concrete form is a \textit{gateway} or proxy that sits between client and servers and inspects traffic at runtime (e.g., validating tool descriptions, filtering tool outputs, and enforcing policy on invocations \cite{mcp-guardian-detection,truefoundry-gateway}). A gateway is the natural enforcement point because it is the one place that sees every call. It counters injection and poisoning at the moment of use rather than relying on the model to resist them.

\textbf{Provenance and signing.} Tool poisoning and supply-chain attacks both exploit the mutability and unverified origin of tool definitions. Signing them at the source addresses the root cause: the Enhanced Tool Definition Interface (ETDI) binds tool definitions to OAuth-backed cryptographic identities with immutable, versioned manifests, so that a tampered or silently-updated tool fails verification \cite{etdi}. Manifest signing and registry provenance generalize this approach \cite{tool-manifest-signing}.

\textbf{Custody as defense.} The key-custody spectrum of Section~\ref{sec:bg-custody} is itself a defense gradient: scoped delegation under EIP-7702, MPC split-authority, and TEE-isolated keys each shrink the blast radius of a compromised agent by constraining what it can sign. Policy-constrained wallets add per-transaction limits and allowlists, so that even a successful injection cannot move an arbitrary value. These directly blunt the signing authority amplifier.

\textbf{Identity.} Against the impersonation surface of Section~\ref{sec:bg-identity}, signed agent cards and mutual TLS authenticate which agent is which, raising the bar for replay and impersonation at the inter-agent layer \cite{a2a-security}.

\subsection{Blockchain-based defenses}

A distinct line of work inverts the paper's direction, using blockchain mechanisms to secure the agent-tool ecosystem itself. Three categories recur. \textit{On-chain registries} record signed tool provenance and revocation as a tamper-evident state. \textit{Attested execution} pairs TEEs with on-chain attestation so that a server can prove what code it ran. \textit{Crypto-economic trust} has a tool or agent operators stake value that is slashable on misbehavior. Decentralized MCP architectures combine several of these \cite{demcp}. These approaches are early and carry open questions (e.g., registry governance, attestation cost, and the circularity of using one trust system to bootstrap another), and we present them as a promising but unsettled direction rather than a solution.

\subsection{What current defenses do not cover}

The honest assessment is that defenses are improving but remain insufficient, and the gaps are structural rather than incidental.

First, \textit{measured effectiveness is low.} A systematic benchmark found that existing protections stop fewer than 30\% of attacks on average \cite{mcpsecbench}, and model-level safety is weaker still. Refusal rates against tool-poisoning fall below 3\% even for safety-tuned models \cite{mcptox}. Alignment alone does not defend the tool layer.

Second, \textit{the core weakness is not fully fixable at the protocol layer.} An agent that acts on the content it consumes will act on malicious content that reaches it. This is a consequence of the design, not a bug to be patched. A vendor characterization of one such report as expected behavior rather than a vulnerability illustrates the point. The boundary between feature and flaw is itself contested \cite{ox-stdio}.

Third, \textit{per-call defenses miss sequence-level composition.} Gateways and policy checks evaluate calls individually, but the sequence-level composition amplifier (Section~\ref{sec:amplifiers}) produces harm from an aggregate of individually permitted calls. No widely-deployed defense reasons over the sequence. This gap, together with the unsolved problem of delegation-chain identity, defines much of the research agenda in Section~\ref{sec:related}.

\section{Related Work and Open Problems}\label{sec:related}

\subsection{Positioning}

Several recent surveys touch on the agent--blockchain intersection, but each addresses a different question than this one.

The closest in subject is the systematization of AI agents for blockchain by Romandini et al. \cite{sok-agents-blockchain}, which surveys how agents \textit{assist} blockchain work: analyzing on-chain data, optimizing transaction strategies, and detecting smart-contract vulnerabilities. Our direction is the opposite. We study the attack surface that opens when agents \textit{act on} Web3 through tool calling. Their agents help secure the chain, while ours are the thing to be secured.

A broader interoperability survey by Alqithami \cite{autonomous-agents-blockchain} reviews 317 works on agent--blockchain integration and contributes a five-part taxonomy of integration patterns and a threat model for agent-driven transactions. Its scope overlaps ours more than the others, but its lens is \textit{standards and execution models}. Ours is the security amplification specific to the execution layer, expressed through the four-amplifier framework and the risk-mapping matrix. We differentiate on analytical framing and the defense-coverage artifact, not on cataloging integration patterns.

A systematization of the MCP security ecosystem \cite{sok-mcp-ecosystem} catalogs MCP threats and defenses without a Web3 focus. Our work can be considered  as extending it with the blockchain execution-layer overlay that makes the attack surface qualitatively different. Table~\ref{tab:benchmarks} situates this survey against the verified MCP-security benchmarks. Each measures attack success, but none carries a Web3 execution-layer lens.

\begin{table}[t]
\caption{MCP-Security Benchmarks and This Survey. Only independently verified benchmarks are listed. This work adds the Web3 execution-layer overlay absent from all prior benchmarks.}
\label{tab:benchmarks}
\centering
\footnotesize
\begin{tabular}{@{}p{1.8cm}p{1.3cm}p{1.0cm}p{2.5cm}@{}}
\toprule
Work & Axis & \# attacks & Web3-specific? \\
\midrule
MCPSecBench & 4 surfaces & 17 types & no; protect. $<$30\% \\
MSB & 3 stages & 12 types & no (ASR 40.71\%) \\
MCPTox & tool descr. & --- & no (72.8\% ASR) \\
\textbf{This survey} & surface $\times$ stage & 9 classes & \textbf{yes (4 amplifiers + risk matrix)} \\
\bottomrule
\end{tabular}
\end{table} A general agent-security systematization \cite{dehghantanha-sok} provides the broader backdrop.

\subsection{Scope boundary}

One adjacent problem is deliberately out of scope. The use of agents as autonomous \textit{generators} of exploits inverts the threat model (the agent is the attacker, not the victim or conduit) and is studied separately \cite{exploit-gen-a1}. We exclude it to keep the usage-to-risk direction sharp.

\subsection{From gaps to directions}

The blank and partial mitigation cells of Table~\ref{tab:matrix}, together with the structural defense gaps of Section~\ref{sec:def} are not loose ends but a structured agenda. Section~\ref{sec:rd} develops the five research directions they imply.

\section{Research Directions}\label{sec:rd}

The preceding analysis is not only descriptive: the risk-mapping matrix (Table~\ref{tab:matrix}) and the defense synthesis (Section~\ref{sec:def}) together expose where the field's hardest unsolved problems lie. Each of the four amplifiers generates a distinct research challenge that current techniques do not address, and a fifth, cross-cutting problem concerns how agent-tool security can be evaluated at all. We frame five directions, each with a technical obstacle, an explanation of why current approaches fall short, and concrete research questions. The directions are grounded in the survey's verified findings rather than offered as open-ended speculation.

\subsection{Semantic integrity of tool descriptions}

\textbf{Obstacle.} Because an MCP tool's natural-language description is supplied to the agent as trusted context (Section~\ref{sec:bg}), the description is the primary protocol-level attack surface (Section~\ref{sec:taxonomy}). The unsolved problem is verifying not who authored a description but whether it is \textit{honest}: whether a tool's declared behavior matches its actual behavior.

\textbf{Why current approaches fall short.} Provenance and signing schemes such as ETDI bind a description to a cryptographic identity, but a validly signed tool can still carry a poisoned description. Signing proves origin, not intent. Model-level alignment is weaker still (refusal rates against tool-poisoning fall below 3\% even for safety-tuned models \cite{mcptox}), and gateway scanning detects surface patterns, not semantic deception. None of these verifies meaning.

\textbf{Research questions.}
\begin{itemize}
\item Can a tool's description be formally checked for conformance against its observed input--output behavior, turning ``description integrity'' into a runtime contract rather than a publish-time signature?
\item What is the detection ceiling for semantic poisoning that uses no known trigger tokens and remains consistent with the tool's nominal function?
\item Can agents be trained or instrumented to treat tool descriptions as untrusted input without losing the usability that makes MCP valuable?
\end{itemize}

\subsection{Sequence-level defense}

\textbf{Obstacle.} The sequence-level composition amplifier (Section~\ref{sec:amplifiers}) produces harm from an aggregate of individually authorized calls (e.g., a permitted read, a permitted computation, and a permitted transfer chaining into a drain or exfiltration pipeline) where no single call violates a policy. The harmful object is the trajectory, not any call within it.

\textbf{Why current approaches fall short.} Gateways and policy engines evaluate calls in isolation by construction. No widely deployed defense reasons over a sequence (Section~\ref{sec:def}). This is the amplifier least covered by current mitigations. It appears across the tool-poisoning, indirect-injection, supply-chain, and preference-manipulation rows of Table~\ref{tab:matrix}, yet every listed mitigation is per-call.

\textbf{Research questions.}
\begin{itemize}
\item Can a stateful monitor flag a dangerous sequence of permitted calls without an unacceptable false-positive rate on benign multi-step tasks?
\item What is the minimal cross-call state required to detect canonical drain and exfiltration patterns?
\item Can per-session capability budgets, for example, a cumulative value-at-risk bound across a trajectory, limit compositional harm without crippling legitimate automation?
\end{itemize}

\subsection{The irreversibility and autonomy gap}

\textbf{Obstacle.} Blockchain settlement is final, and agents act continuously without per-step human review. The window between a successful injection and an irreversible on-chain action is therefore bounded only by the agent's own latency, not by a human's. Two amplifiers (irreversibility and continuous autonomy) compound here: there is no human-speed window in which to intervene, and once a transaction settles, after-the-fact detection is moot.

\textbf{Why current approaches fall short.} Human-in-the-loop confirmation does not scale to continuous autonomous operation, and conventional incident response assumes an observable, reversible window that on-chain execution does not provide. The defenses of Section~\ref{sec:def} reduce the probability of compromise but do not address what happens in the interval between compromise and irreversible settlement.

\textbf{Research questions.}
\begin{itemize}
	\item Can high-stakes actions be reversibly staged (e.g., through commit--reveal, time-locks, or challenge periods) so that an irreversible action becomes interruptible without destroying agent utility?
\item Which on-chain primitives (escrow, optimistic delay, circuit breakers) most effectively convert irreversible actions into recoverable ones for autonomous agents?
\item How should an agent calibrate when to demand human confirmation as a function of value-at-risk and the trust level of the triggering input?
\end{itemize}

\subsection{Delegation-chain identity and accountability}

\textbf{Obstacle.} When an agent acts on behalf of a user who, in turn, acts for an organization, both \textit{authority} (what may this chain legitimately sign?) and \textit{accountability} (who is responsible for a given action?) are poorly modeled. This is the matrix's least-covered row (Table~\ref{tab:matrix}, identity): an impersonated or over-privileged agent in such a chain can exercise signing authority it was never meant to hold.

\textbf{Why current approaches fall short.} Signed Agent Cards and mutual TLS authenticate a single hop, not a chain of delegations. Scoped on-chain delegation, such as EIP-7702, governs what an account may sign but is not bound to the off-chain chain of principals behind the agent, so a compromised intermediate agent can act within its on-chain grant while exceeding its actual mandate.

\textbf{Research questions.}
\begin{itemize}
\item Can delegation be made verifiable end-to-end: a cryptographic chain linking a human principal through intermediary agents to an on-chain action?
\item How should scoped signing authority be bound to a verified delegation chain so that a compromised intermediate agent cannot exceed its delegated grant?
\item What revocation model is appropriate when one link in a live delegation chain is found to be compromised?
\end{itemize}

\subsection{Evaluation and assurance under non-determinism}

\textbf{Obstacle.} Agent behavior is non-deterministic, which frustrates the test-and-certify model that secures conventional software. At present, the field can benchmark the average failure rate of an agent-tool deployment, but cannot certify its safety.

\textbf{Why current approaches fall short.} Existing benchmarks measure average attack-success rate on fixed attack sets, protections stop fewer than 30\% of attacks on average \cite{mcpsecbench}, but they neither bound worst-case behavior nor generalize to unseen attacks. Compounding this, the boundary between a vulnerability and intended behavior is itself contested: when an agent acting on consumed content is exploited, vendors may classify the report as expected behavior rather than a flaw \cite{ox-stdio}, leaving disclosure and response norms undefined.

\textbf{Research questions.}
\begin{itemize}
	\item What would a certification regime (as opposed to a benchmark) for agent-tool security look like: assurance cases, bounded guarantees, or worst-case rather than average-case metrics?
\item Can a security metric be defined that bounds the worst-case behavior of a non-deterministic agent over a space of adversarial inputs?
\item How should disclosure and incident-response norms be standardized for an ecosystem in which the exploited behavior is frequently the intended behavior?
\end{itemize}

\subsection{Limitations}
Three caveats bound the claims above. The survey is scoped to the usage-to-risk direction and treats defenses only insofar as they map to that surface. The risk-mapping matrix (Table~\ref{tab:matrix}) assigns qualitative severity rather than measured scores, and the CVSS values we cite are vendor-reported, with the one quantitative incident figure (Section~\ref{sec:mechanism}) explicitly hedged. Finally, the MCP-security literature moves faster than any survey can track, so we anchor every cell to a confirmed identifier or documented incident, keeping the scaffolding usable as rows are superseded.

\section{Conclusion}

As agents move from reading to acting, the tools that grant them authority have become the attack surface. This survey has argued that when that authority is exercised on Web3, the blockchain execution layer turns the recoverable failures of generic agent security into standing, irreversible loss, through four amplifiers. The attack-surface taxonomy and the risk-mapping matrix organize this argument into durable artifacts: the taxonomy locates the known attack classes, and the matrix ties each to its amplified Web3 impact and to the defenses that do, and do not, cover it. Those defenses remain insufficient, and their gaps are structural: semantic tool integrity, sequence-level reasoning, delegation-chain identity, and safe interruption of irreversible action. The ecosystem is being built now, while these problems remain open.

\clearpage
\bibliography{references}

\end{document}